\documentclass[aps,prl,twocolumn,superscriptaddress]{revtex4}

\usepackage{graphicx}

\usepackage{amsmath}

\newcommand{\bv}[1]{{\bf #1}}

\begin{document}
 
\title{Spectral weight transfer in a disorder-broadened Landau level}

\author{Chenggang Zhou} \affiliation{Center for Nanophase Materials
Science, Oak Ridge National Laboratory, P.O. Box 2008, MS 6493, Oak
Ridge Tennessee, 37831-6493 USA } \author{Mona Berciu}
\affiliation{Department of Physics and Astronomy, University of
British Columbia, Vancouver B.C. V6T 1Z1, Canada}

\maketitle

{\bf In the absence of disorder, the degeneracy of a Landau level (LL)
is $N=BA/\phi_0$, where $B$ is the magnetic field, $A$ is the area of
the sample and  $\phi_0=h/e$ is the magnetic flux quantum. With
disorder, localized states appear at the top and bottom of the
broadened LL, while states in the center of the LL (the critical
region) remain delocalized. This well-known phenomenology is
sufficient to explain most aspects of the Integer Quantum Hall Effect
(IQHE)~\cite{Klitzing80}. One unnoticed issue is where  the new states 
appear as the magnetic field is
increased. Here we demonstrate that they appear predominantly inside
the critical region. This leads to a certain ``spectral ordering'' of
the localized states that explains the stripes observed in
measurements of the local inverse compressibility~\cite{Ilani04,
Martin04}, of two-terminal conductance~\cite{Cobden99},  and of Hall and
longitudinal resistances~\cite{Jouault07} without invoking interactions as 
done in previous work~\cite{Pereira06, Struck06, Sohrmann06}.  }

The spectrum and eigenstates of a disorder-broadened LL can be studied
with the well-established approach of diagonalizing the single
electron Hamiltonian
\begin{equation}
\label{eq1}
{\cal H} = {(2m_e)^{-1}} \left[-i\hbar \nabla + e \bv{A}({\bv r})
\right]^2 + V(\bv r),
\end{equation} 
where we choose $\bv{A}({\bv r}) = (0,Bx)$, $V(\bv{r})$ is the
disorder, and periodic boundary conditions (PBC) are applied to a
system of area $A=L\times L$. Properties of single particle states,
such as the localization length, are calculated for each eigenstate
and then averaged over disorder realizations.  Many-body wavefunctions
(Slater determinants) are constructed from these single-electron
states. Usually, in theoretical studies the magnetic field is kept
fixed at a value $B = N\phi_0/L^2 $, where $N$ is an integer defining
the dimension of the LL subspace, and one sweeps the electron density
$n_e$ or, equivalently, the filling factor $\nu = n_e A/N$, by
adjusting the Fermi energy $E_F$.

Since the experiments mentioned above investigate behavior of various
quantities in the $(n_e, B)$ plane, we need to understand how the
spectrum changes when $B$ is also tuned. Given the constraint that an
integer number of fluxes must penetrate the sample, $B$ can only
change in discrete steps of $\phi_0/L^2$. Therefore, we ask the
following question: How do single electronic wavefunctions evolve when
one more magnetic flux is inserted?

Let $|i,N\rangle$, $1\le i \le N$ be the eigenstates of a
spin-polarized LL corresponding to a given disorder $V(\bv r)$ and a
magnetic field $B=N\phi_0/L^2$. The states are ordered by their
energies $E_1 <E_2< ...< E_N$ (accidental degeneracies can be lifted
with minute changes in $V(\bv r)$).

To see how the wavefunctions evolve when $B$ increases, we calculate
their disorder-averaged overlaps:
\begin{equation}
	\overline{D}_N(i,j) = \overline{\left| \left<i,N | j, N+1\right>
	\right|^2},
	\label{eq2}
\end{equation}
where $1\leq i \leq N $ and $1 \leq j \leq N+1 $ label two eigenstates
of the Hamiltonian~(\ref{eq1}) with the same disorder potential but
different magnetic fields. The overline indicates a disorder
average. In the results presented here we typically average over 1000
disorder realizations, and show results only for the spin-polarized
lowest LL. Similar results are expected in higher LLs. The disorder
potential $V(\bv{r})$ is modeled as a sum of many short range,
randomly placed Gaussian scatterers. We show 4 sets of data, for
$L= 50, 128, 200, 750$ nm,  $N=50, 128, 200, 550$ and therefore
$B = 1.654$~T for the first three data sets, and 2.022~T for the last data set.

\begin{figure}[t]
\includegraphics[width=\columnwidth]{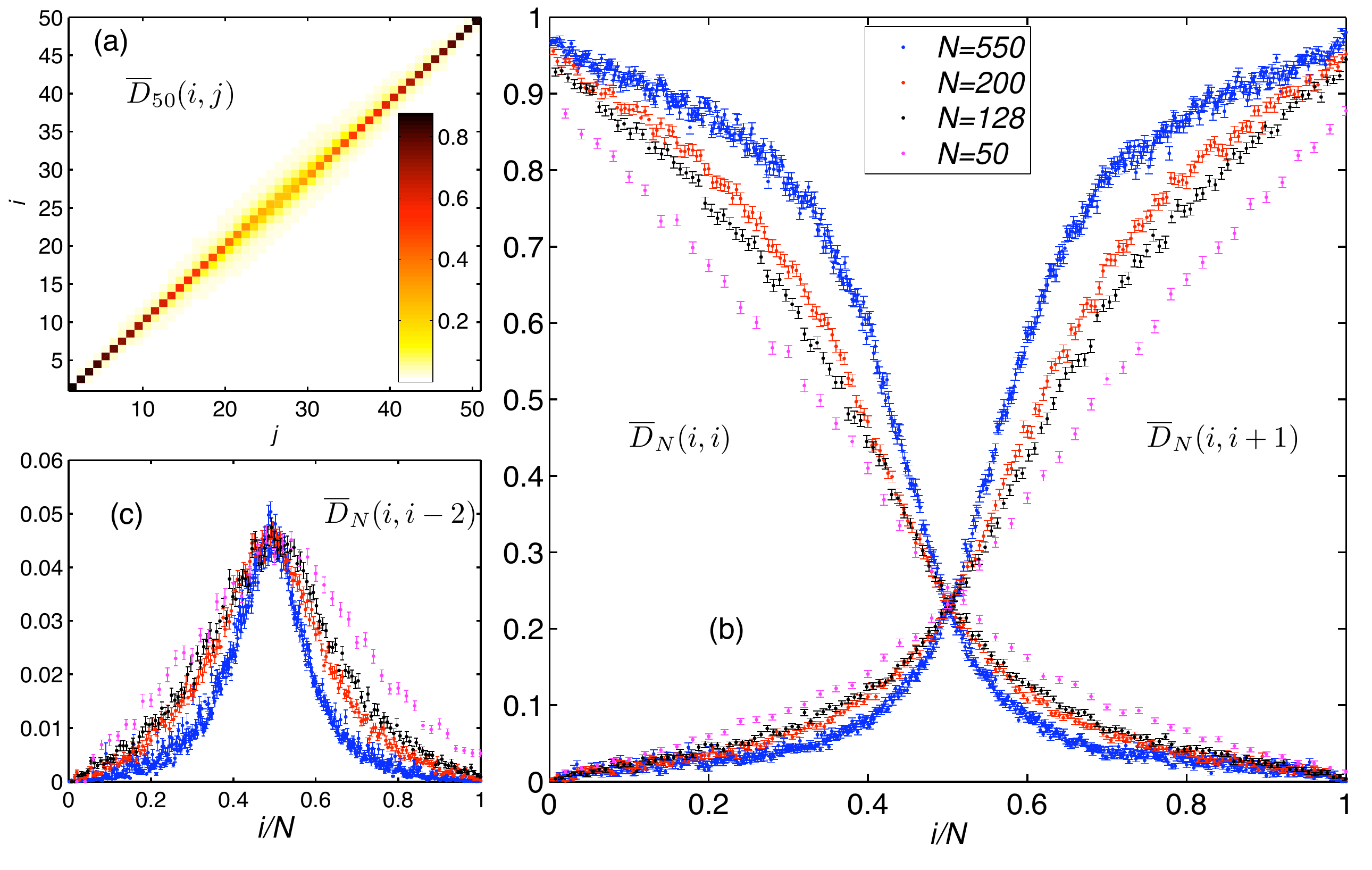}
\caption{Overlap between eigenstates with an additional
magnetic flux quantum. (a) Plot of the 50 by 51 overlap matrix $\overline{D}_{50}$,
defined in Eq. (\ref{eq2}). Only elements near the diagonal are visible.  
(b) $\overline{D}_{N}(i,i)$ and $\overline{D}_{N}(i,i+1)$
vs. filling factor $i/N$ for 4 different sizes $N=50, 128, 200,
550$. The transition region narrows towards half filling as $N$
increases.  (c) Typical off-diagonal overlap matrix elements
$\overline{D}_{N}(i,i-2)$ vs. $i/N$, for the same values of $N$. These
are much smaller than the main diagonal elements shown in (b). The
peak narrows as $N$ increases.}
\label{fig1}
\end{figure}

 The $N$ by $N+1$ matrix $\overline{D}_N(i,j)$ is almost zero
everywhere except near its diagonal, as shown in
Fig.~\ref{fig1}(a). Focusing on this region in Fig~\ref{fig1}(b), we
see that $\overline{D}_N(i,i)$ decreases from near unity to near zero
as $i$ increases from 1 to $N$, while $\overline{D}_N(i,i+1)$ is
almost the mirror image of $\overline{D}_N(i,i)$. 
They intersect at $j/N\approx 1/2$, i.e. half filling, where they seem to have a universal
value, which is independent of $N$ and $B$. These elements
change most rapidly in a region near half-filling, which becomes
narrower as $N$ increases.  Other matrix elements, {\em e.g.}
$\overline{D}_N(i,i-2)$ shown in Fig.~\ref{fig1}(c), exhibit a very
small peak in this narrow region which we identify as the critical
region.

Since the overlap of Eq.~(\ref{eq2}) measures the similarity of
eigenstates, these results show that the new state created when $B$
increases by one flux quantum appears predominantly in the center of
the disorder-broadened LL. Localized states at the bottom of the LL
are little affected and keep the same spectral ordering leading to
large $\overline{D}_N(i,i)$ overlaps. Localized states at the top of
the LL also keep their spectral ordering but shifted upward by 1, to
account for the new eigenstate created in the critical region. This
explains why $\overline{D}_N(i,i+1)$ is here close to unity.

The appearance of the new state amongst the delocalized states is not
surprising since such states can enclose a large area, with sufficient
of the additional magnetic flux going through, and therefore the
effects of the small $\delta B = \phi_0/L^2$ increase are not perturbative even
for a large $B$. By contrast, for localized states the effect of the
additional flux is perturbatively small, leading only to slight
spatial deformations of the wavefunctions.

This conclusion can also be reached using well-known results for the
Hofstadter butterfly. To map into these, we use copies of the $L\times
L$ system to tile the infinite plane, so that the disorder $V(\bv r)$
becomes periodic with period $L$. The resulting Hofstadter problem has
as magnetic unit cell the $L\times L$ area, thus it corresponds to
$BL^2/\phi_0 = q/p= N$. The eigenstates of Hamiltonian (\ref{eq1}) are
now magnetic Bloch waves $\psi_{i,\bf k}(\bv r) = e^{-i\bv{k}\cdot
\bv{r}} u_{i,\bv{k}} (\bv r)$. The integer $i$ labels the $q=N$
magnetic Bloch bands (MBBs) originated from an LL. The functions
$u_{i,{\bf k}}(\bv r)$ satisfy generalized PBC~\cite{Thouless82}. In
effect, each of the $N$ eigenstates of an LL of the finite-size
$L\times L$ system has evolved into an MBB of the Hofstadter problem
(the former are the ${\bv k} = 0$ states of the latter).  As a result,
we can associate to each eigenstate of the finite-size system the
Chern number $\sigma_i$ of the corresponding
MBB~\cite{Thouless82}. $\sigma_i$ is well defined for each MBB,
because energy bands $E_i(\bv k)$ and $E_{i+1}(\bv k)$ can only touch
at discrete $\bv k$ points, and small changes in $V({\bv r})$ can
remove such degeneracies, as implicitly assumed when $\sigma_i$ is
calculated in Refs.~\cite{Huo92,Yang97}.

Thouless showed that localized states have zero Chern
numbers~\cite{Thouless84}. This is easy to understand, since localized
states are rather insensitive to changes in the boundary conditions
used to calculate the Chern number~\cite{Thouless82}. This is verified
by numerical calculations and finite size scaling
analysis~\cite{Huo92, Yang97} showing that the non-zero Chern numbers
are distributed near the center of the LL. The distribution fits the
scaling theory of IQHE~\cite{Pruisken87, Huckestein90} with the
correct localization exponent $\nu = 2.34\pm 0.04$~\cite{Huckestein90,
Wei88}.

When the magnetic field is increased by $\phi_0/L^2$, {\em i.e.} $N
\rightarrow N+1$, a new MBB must appear in the spectrum generated from
each LL, thus one of the original MBBs must split into two. We now
argue that only an MBB with a non-zero Chern number can do this, in
other words the new MBB (new state) appears in the critical region.

A simple proof is obtained from combining Thouless {\em et al.}'s
famous proportionality between the conductance of a MBB and its Chern
number~\cite{Thouless82}, and St\v{r}eda's formula~\cite{Streda}
linking the conductance to the change in the density of states with
changing $B$. This gives an expression for the Chern number:
\begin{equation}
\sigma_i = he^{-1}{\partial N_i(B) / \partial B },
\end{equation}
where $N_i(B)$ is the density of states in the $i$th MBB. If this
corresponds to a localized state, then $\sigma_i=0$~\cite{Thouless84}
and this MBB cannot be the origin of the new state since its density
of states stays unchanged as $B$ varies. The new electronic state in
this LL must therefore originate from subbands having non-zero Chern
numbers.

Another proof for the above result is obtained from the semi-classical
theory of Chang and Niu~\cite{Niu95, Niu85}.  If $\delta B=\phi_0/L^2$
supplies the additional flux quantum, the quantization condition of
hyperorbits in the MBB reads
\begin{equation}
\label{eq3}
{\hbar \over 2e\delta B} \oint_{C_m} (\bv{k} \times d\bv{k})\cdot
\bv{z} + \Gamma_i(C_m) = 2\pi (m+ 1/2),
\end{equation}
where $m$ is an integer and $\Gamma_i(C_m)$ is the contour
integral over an effective gauge field ${\cal A}_i$.  Chang and Niu
obtained the entire hierarchical structure of the Hofstadter butterfly
by approximating the integral in Eq.~(\ref{eq3}) with the area of the
magnetic Brillouin zone, and replacing $\Gamma_i(C_m)$ with the
Chern number.  If $\sigma_i = 0$, the localized wavefunction
$\psi_i(\bv{k})$ can be expanded as an absolutely convergent sum of
Wannier functions~\cite{Thouless84}, and the curvature of ${\cal A}_i$
vanishes identically. Thus, $\Gamma_i(C_m)$ indeed vanishes for
any localized MBB regardless of the shape of the hyperorbit
$C_m$. Also, since our magnetic Brillouin zone is $[-\pi/L, \pi
/L)\times [-\pi/L, \pi/ L) $ by construction, the l.h.s. of
Eq.~(\ref{eq3}) is found to be less or equal to $\pi$. It follows that $m = 0$ is
the unique possibility for the MBB of a localized state, i.e. such an MBB
does not split into multiple MBBs when the magnetic field increases by
$\delta B$~\cite{Niu95, Niu85}. Note that both these arguments are
only valid if gaps between neighboring MBBs remain open as $B$
increases by $\delta B$. As already argued, this is expected to be
typically the case, since accidental degeneracies closing the gap can
be removed by small changes in the disorder $V(\bf r)$.

\begin{figure}[t]
\includegraphics[angle = 90, width=\columnwidth]{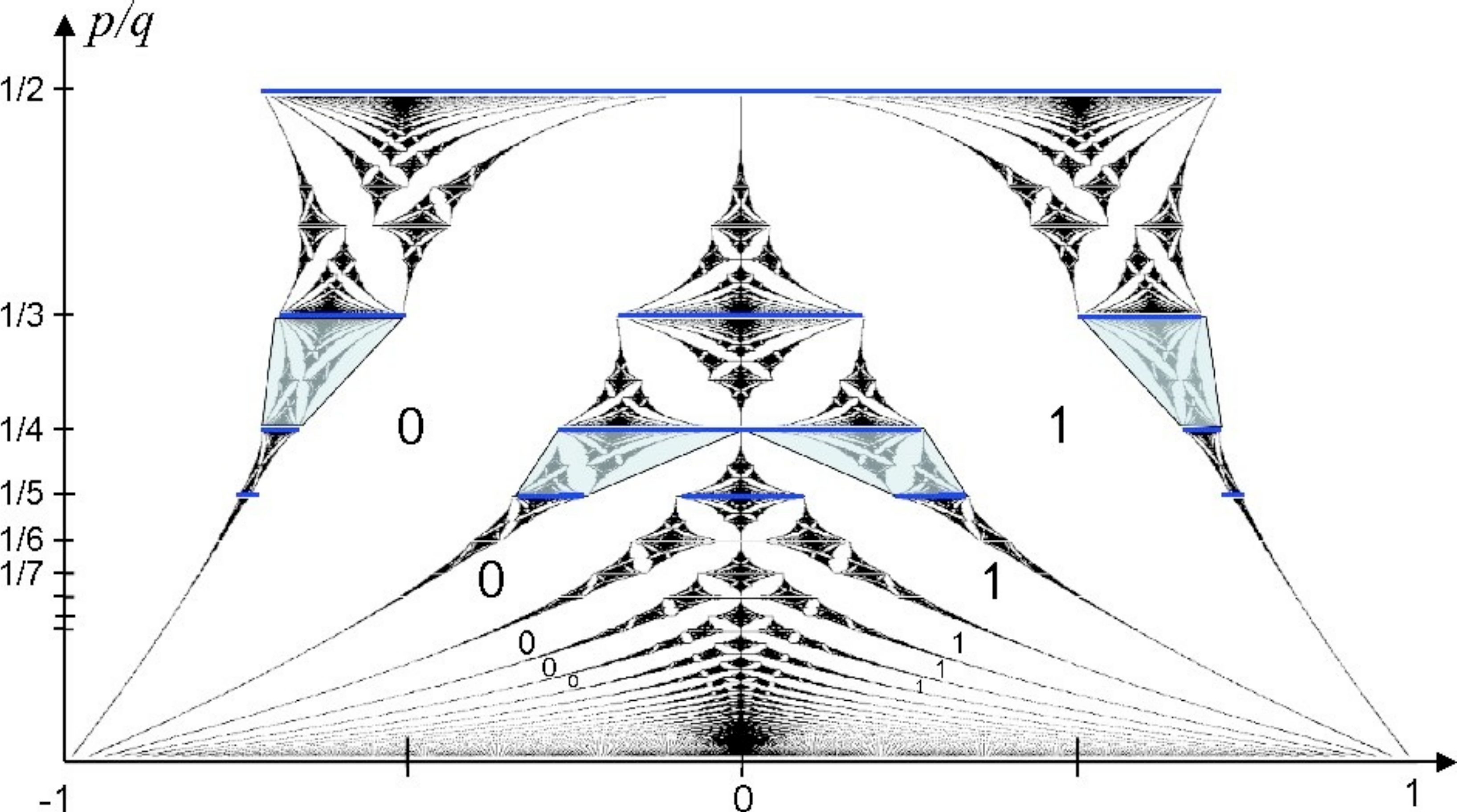}
\caption{Lower half of a Hofstadter butterfly. The subbands for $p=1,
q=N$ are marked by thick blue lines. The Hall conductances in units of
$e^2/h$ are given for the main gaps, which never close as $B\sim q/p$
increases. The shaded blocks are typical self-similar spectra
generated by an MBB when an additional flux is inserted ($N\rightarrow
N+1$). The new spectral weight as $B$ increases always appears in the
center of the LL.}
\label{fig3}
\end{figure}

A nice illustration of this property is given by the very simple
``disorder'' potential $V({\bf r})\sim \cos(2\pi x/L) + \cos (2\pi
y/L)$. Of course, this leads to the well-known Hofstadter butterfly
spectrum, whose lower half is shown in Fig.~\ref{fig3}. In accordance
with our discussion, we are only interested in magnetic fields $B$
corresponding to $q/p=N$, for large $N$, marked by thick lines in the
figure. As expected, there are $N$ MBBs (for even $N$, the two central
MBBs just touch). When $N\rightarrow N+1$, a new MBB is spawned from
the central MBB(s), which is the only one with a non-zero Chern
number. Indeed, if $N$ is odd, the central subband evolves into two
subbands, whereas if $N$ is even, a new subband grows out in the
center. The outside MBBs have zero Chern numbers, and indeed
correspond to states localized about the bottom/top of this
``disorder'' potential. As argued above, the spectrum of a general
disorder potential also has these properties, except that typically
there are several MBBs with non-zero Chern numbers, one of which will
generate the new state when $B$ increases by one magnetic flux.

To summarize, all these numerical and theoretical arguments prove that
the new states generated in a disorder-broadened LL when the magnetic
field increases appear predominantly in the critical region. After a state is 
expelled to the upper (lower) localized
regions as $B$ increases, its order from the top (bottom) of the LL
remains essentially fixed.

\begin{figure}[t]
\includegraphics[width=0.9\columnwidth]{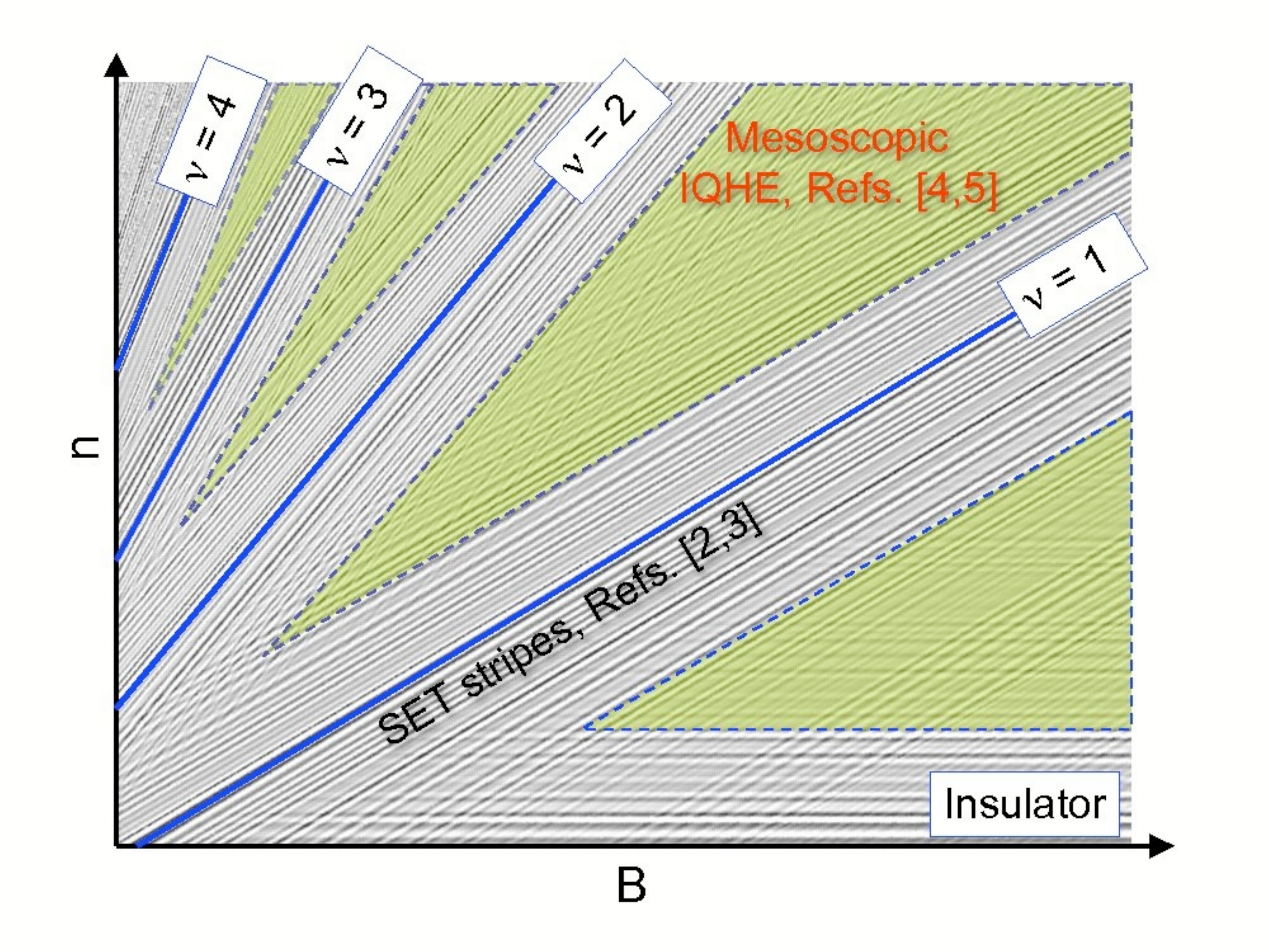}
\caption{A schematic composite picture (computer-generated) of the
  stripes observed experimentally in Refs.~\cite{Ilani04,
  Martin04,Cobden99,Jouault07}. }
\label{fig2}
\end{figure}

This spectral ordering is the main ingredient needed for understanding
the results of recent single-electron transistor (SET)
measurements~\cite{Yoo97,Zhitenev00} that investigate the charge
distribution of localized electronic states in two-dimensional
electron systems~\cite{Ilani04, Martin04}, as well as of measurements
of mesoscopic fluctuations of two-terminal
conductances~\cite{Cobden99} and of Hall and longitudinal
resistances~\cite{Jouault07}. When plotted in the $(n_e, B)$ plane,
the maxima in these quantities are found to track straight lines with
certain quantized values for their slopes, as described below. This
suggests that such ``stripes'' are an intrinsic aspect of IQHE
phenomenology.  In fact, SET and transport experiments are strikingly
complementary to each other. When their results are put together, as
schematically shown in Fig.~\ref{fig2}, we get a complete picture of
these stripes: States belonging to the $n$th LL are located between
the straight lines $n_e = nB/\phi_0$ and $n_e = (n+1) B/\phi_0$. In
the upper half of this LL, stripes are found to be parallel with
$n_e=(n+1) B/\phi_0$, while in the lower half, stripes are parallel to
$n_e = n B/\phi_0$. Near the center of the LL, stripes of both slopes
are visible and can cross each other. Ref.~\cite{Ilani04} images the
stripes close to the LL edges, while Refs.~\cite{Cobden99,Jouault07}
image the stripes near LL centers.

So far it has been unanimously agreed that these stripes are
signatures of Coulomb-blockade physics in the localized
states~\cite{Pereira06, Struck06, Sohrmann06}. We now argue that the
main reason for these stripes' appearance is in fact the spectral
ordering discussed above, which is a single-electron
effect. Interactions do play a role through screening, as discussed
below, but it is very much a secondary one.

We begin our discussion with the SET results that measure the ``local
inverse compressibility" $d\mu / dn_e$, which is a local DC response
function dominated by localized states located under the SET
tip. Consider the evolution of a maximum due to one such localized
state, for example one that is found near the bottom of the $n$th
LL. If the magnetic field is increased by $\delta B = \phi_0/L^2$, $n$
new states appear near the centers of the lower $n$ LLs (counting from
$n=0$). As a result, in order to bring the Fermi level back to this
particular state so as to see the same maximum, $n_e$ must be
increased by $\delta n_e = n/L^2$. Thus, the maximum moves along a
line of slope $\delta n_e/\delta B = n/\phi_0$. For a state at the top
of the $n$th LL, however, the density change must be $\delta n_e =
(n+1)/L^2$, since the spectral position of this state is also shifted
upwards by the new state appearing near the center of the $n$th LL
itself. Thus, maxima due to these states will have a slope of
$(n+1)/\phi_0$, precisely as seen in experiments.

It is worth noting that Fig 4. of Ref.~\cite{Sohrmann06} shows that
even if the Coulomb interaction is turned off, stripes do appear with
essentially the right slopes. The authors argue that these are not in
agreement with experiment because the region occupied by them
increases with $B$, whereas in experiments one sees a roughly constant
number of maxima, as sketched in Fig. \ref{fig2}. Addition of Coulomb
interactions fixes this problem, but this is because of screening:
their results show that the stronger the interaction, the more
effective the screening, the fewer states (maxima) are seen. We
therefore argue that Coulomb interactions (screening) have the
secondary role of limiting the number of localized states ``visible''
to the tip, but the stripes' slopes are determined purely by
single-electron physics.

Note that this explanation relies essentially on the fact that
localized states tend to keep their spectral order with respect to the
top or bottom of the LL. Of course, states localized about the same
minimum or maximum in the disorder landscape do keep their relative
spectral ordering, but it is possible that the energies of states
localized in different spatial regions might cross each other as $B$
varies. Such events must be rare, as our simulations in
Fig.~\ref{fig1} show; in fact, we find that $\overline{D}_N(i,i)$ and
$\overline{D}_N(i,i+1)$ get closer to 1 in the relevant interval when
$N$ increases! However, if such a rare crossing does take place for
one of the states under the SET tip, the maximum will shift by $\delta
n_e = \pm 1/L^2$ at the $B$ value where the crossing occurs, after
which the stripe resumes with the correct slope. Such jumps would be
rather impossible to measure.

The stripes in the transport measurements have the same origin. Here,
the mesoscopic fluctuations are caused by electronic states that
mediate the charge transport across the Hall bar, as shown in
Ref.~\cite{Zhou05, Berciu05}. For example, the fluctuations in the two-terminal
conductance measured in Ref.~\cite{Cobden99} are due to Jain-Kivelson
tunneling~\cite{Jain88} through electronic states located in the
central region of the sample. To see the same resonance, the Fermi
level must be tuned to match the energy of the state mediating the
tunneling, so one expects to see the maxima following lines of
quantized slopes in the $(n_e, B)$ plane for the same reasons given
above. However, unlike for SET measurements, the Fermi level is now near 
the center of the LL, where the transition between quantum Hall
plateaus occurs.  When an additional $\phi_0$ is inserted, there may
be either $n$ or $n+1$ new states below the Fermi level. As a result, one
expects to see stripes with both slopes. Occasional crossings of
stripes is also expected, when the 
states mediating their tunneling are more than a coherence length
$L_\phi$ apart, i.e. there is no quantum interference between these
two tunneling events.
 
These arguments also explain the stripes observed in SET measurements
~\cite{Martin04} in the fractional quantum Hall effect
(FQHE)~\cite{Tsui82}. It is well known that FQHE can be explained as
IQHE of quasi-particles~\cite{Laughlin83, Haldane83, Jain89}. Our
explanation of the stripes in the IQHE regime is equally applicable to
the FQHE regime, except with electrons replaced by these
quasi-particles.

In conclusion, we claim that the spectral ordering of states within a
LL, demonstrated in the first part of this work, is the main
ingredient needed to understand the stripes observed in SET and transport
measurements. Unlike other authors who identify them as 
effects of interactions, we find that single-electron physics suffices
to explain, in a unified manner, all these observations. This is
gratifying, given the 
long-standing success of the non-interacting electron 
approximation in describing all other aspects of IQHE physics.

\acknowledgments
Acknowledgments: We  thank B. Jouault and X.-G. Zhang for many
stimulating discussions and insightful opinions.  
CZ acknowledges support from the Center for Nanophase Materials
Sciences, sponsored at Oak Ridge National Laboratory by the Division
of Scientific User Facilities, U.S. Department of Energy.  MB
acknowledges support from the Sloan Foundation, CIfAR Nanoelectronics
and NSERC.

Competing interests statement: 
The authors declare no competing financial interests.

\end{document}